\documentstyle[aps,twocolumn]{revtex}
\input psfig
\begin{document}
\title{Aharonov-Bohm-Type Oscillations of Thermopower in a Quantum
Dot Ring Geometry}
\author{Ya.~M. Blanter$^{a,b}$, C. Bruder$^c$, Rosario Fazio$^d$, and 
	Herbert Schoeller$^c$}
\address{\mbox{$^a$ Institut f\"ur Theorie der Kondensierten Materie,
	Universit\"at Karlsruhe, 76128 Karlsruhe, Germany}
\mbox{$^b$ Theoretical Physics, Moscow Institute for Steel 
	and Alloys, Leninskii pr. 4, 117936 Moscow, Russia}
\mbox{$^c$ Institut f\"ur Theoretische Festk\"orperphysik,
	Universit\"at Karlsruhe, 76128 Karlsruhe, Germany}
\mbox{$^d$ Istituto di Fisica, Universit\`a di Catania, 
	viale A. Doria 6, 95128 Catania, Italy}}
\date{\today}
\maketitle 
\tighten

\begin{abstract}
We investigate Aharonov-Bohm-type oscillations of the thermopower of
a quantum dot embedded in a ring for the case when the interaction 
between electrons can be neglected. The thermopower is shown to be 
strongly flux dependent, and typically the amplitude of oscillations 
exceeds the background value. It is also shown to be essentially
dependent on the phase of the scattering matrix which is determined by the
experimental geometry and is not known in the given experiment. Two
procedures to compare theory and experiment are proposed. 
\end{abstract}

The interest in phase-sensitive measurements has increased recently
due to a series of beautiful experiments by Yacoby {\it et al.} 
\cite{Yacoby} and Schuster {\it et al.} \cite{Schuster} in which the 
electron phase shift due to transmission through a quantum dot has
been directly measured. These experiments led to a number of theoretical
papers \cite{Buett,Hack,BFS,Oreg} in which the phase dependence
of the conductance in the presence of a quantum dot was investigated.

Below we present a theoretical investigation of phase-sensitive
effects in the thermopower. The thermopower of a system of electrons is 
extremely sensitive to the energy dependence of the density of states 
\cite{Abr}. As examples, we mention the behavior of the thermopower in 
Kondo systems \cite{Kondo} and the sensitivity to the band
structure \cite{etp}. In both cases the thermopower exhibits
singularities and can be experimentally used to detect
the corresponding effects. Shubnikov-de-Haas-type oscillations of
the thermopower are a similar phenomenon \cite{osc}: the amplitude of 
oscillations exceeds the mean value of the thermopower, and, as a 
consequence, the thermoelectric effect changes sign as a
function of the applied magnetic field. 

In the present paper we investigate the thermopower of a particular
mesoscopic system, corresponding to the experiment \cite{Yacoby} -- a
quantum dot embedded in an Aharonov-Bohm (AB) ring. While the 
flux dependence of the conductance was studied, e.g. in
Ref. \onlinecite{BFS}, in both the linear and nonlinear regime, for
interacting and non-interacting electrons, we concentrate below on the
case of linear transport and consider non-interacting electrons only. 
We find that the thermopower exhibits AB-type oscillations; in contrast to
the conductivity, typically the amplitude of these oscillations
exceeds the mean value of the thermopower, causing a sign change of
the thermoelectric effect vs. magnetic field. Moreover, we show that
the shape of these oscillations essentially depends on the phase of
the scattering matrix. This phase is an individual characteristic of a
given system, and is determined by its microscopic details. It remains
unknown in the given experiment, complicating direct comparison with
the theory. We suggest two ways of overcoming this difficulty. The first
one is to vary this phase in the experiment in the spirit of
Refs. \onlinecite{Yacoby,Schuster}; another way is to consider it as a
random variable. Statistical fluctuations of the thermopower with
respect to this variable exceed the mean value. Our calculations can
be easily generalized  to arbitrary scattering geometries, for which
we expect similar results.  

Recent progress in the investigation of the thermopower of
mesoscopic systems both on the experimental 
\cite{Gallagher,Mol,Staring,Dzurak,Wyss} as well as on the theoretical
\cite{SI,Esp,Lesovik,Beenakker,Houten} 
side allows us to hope that experimental 
studies of the thermopower in this system will soon be available. 

\begin{figure}
\centerline{\psfig{figure=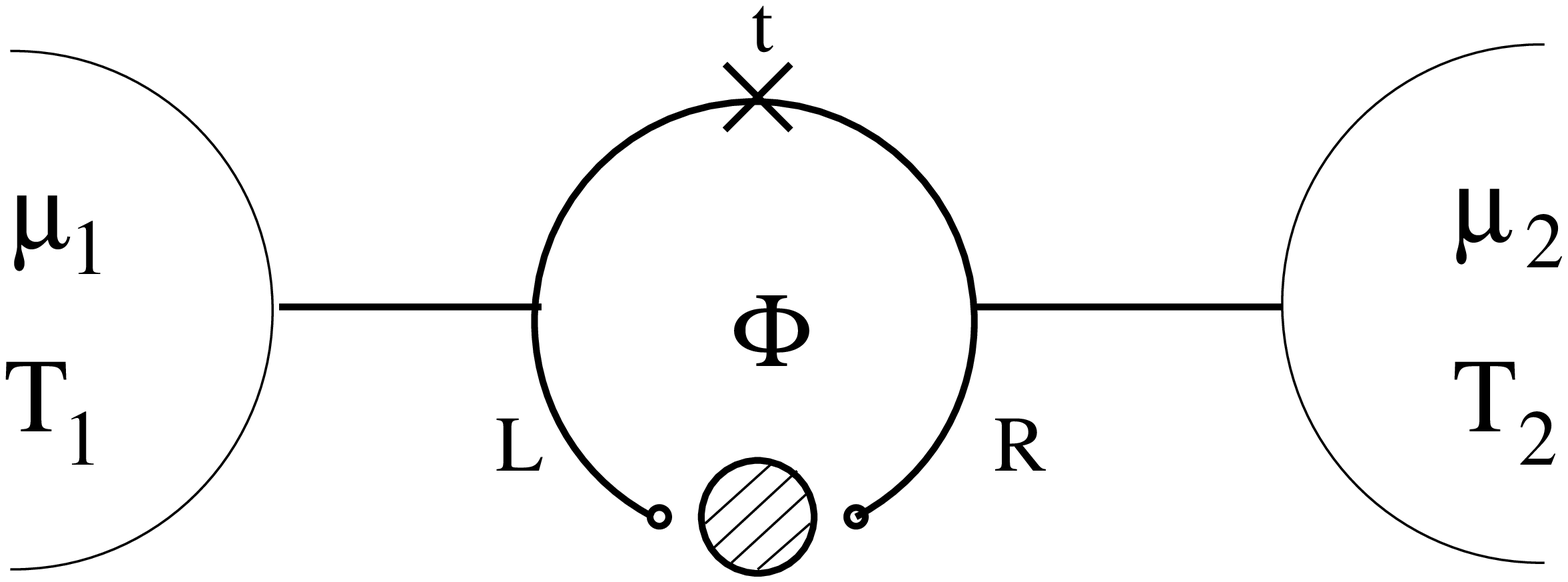,width=8cm}}
\vspace{0.3cm}
\caption{Geometry of the ring connected to reservoirs 1 and 2.
The ring is connected to a quantum dot via high tunneling barriers. 
$\Phi$ is the flux penetrating the ring.}
\label{fig1}
\end{figure}

We consider the two-terminal configuration shown in Fig.~\ref{fig1}.

In the absence of a magnetic field, the upper arm is characterized by the 
scattering matrix 
\begin{eqnarray} \label{scatt1}
\hat T_{up} = \exp(i\theta) \left( \begin{array}{lr} r_1 & t_1\\
t_1' & r_1' \end{array} \right)\; . 
\end{eqnarray}
We now specialize to the case of good transmission through the
upper arm, and consequently choose $t_1 = t_1' = 1$ and
$r_1 = r_1' = 0$. The quantum dot with one single-electron level \cite{foot1}
is described by the scattering matrix 
\begin{eqnarray} \label{scatt2}
\hat T_{dot} = {\exp(i\theta')\over E - \epsilon + i\Gamma}
 \left( \begin{array}{lr}
i\Gamma & -E-\epsilon \\
-E-\epsilon & i\Gamma\\
\end{array} \right)\; .
\end{eqnarray}
Here, $\Gamma$, $E$ and $\epsilon$ are the tunneling rate through the
dot, the electron energy, and the position of the level. The
latter is controlled by an external gate voltage. Only the difference 
between the phases $\theta$ and
$\theta'$ matters, and therefore we put $\theta' = 0$. Then the phase
$\theta$ is acquired by motion along the ring: $\theta = kL + \delta
\theta$, with $k$, $L$, and $\delta \theta$ being the wavenumber, the
ring circumference, and the phase shift in the quantum dot,
respectively, i.e. this phase is a geometrical characteristic of the
system. Furthermore, we assume that the ring is penetrated
by the magnetic flux $\Phi$. 

The transmission coefficient $t(E)$ of
the whole structure was calculated by Gefen, Imry, and Azbel
\cite{GIA}; for our particular scattering matrices (\ref{scatt1}),
(\ref{scatt2}) we obtain:
\begin{equation} \label{scatcoeff}
 t(E) = 4{4(\Delta E)^2 + 4 \Delta E \Gamma \cos \theta \cos \phi
+ \Gamma^2 \cos \theta \over
\lambda_1 (\Delta E)^2 + \lambda_2 \Delta E \Gamma +
\lambda_3 \Gamma^2} \; .
\end{equation}

Here, $\Delta E = E - \epsilon$, $\phi = 2 \pi \Phi/\Phi_0 + \theta$,
$\Phi_0=hc/e$ is the flux quantum, and the quantities $\lambda_i$ are
given by 
\begin{eqnarray*}
\lambda_1 & = & 16 + 9\cos^2 \theta\; ; \\
\lambda_2 & = & \cos \theta (10\cos\phi - 6\sin\theta)\; ; \\
\lambda_3 & = & 1 + \cos^2 \phi + 3\cos^2\theta\; .
\end{eqnarray*}

The ring is connected to two reservoirs, and the current is given by
the usual expression
\begin{equation} \label{curr}
I = (e/2\pi) \int t(E) (f_L - f_R) dE\; ,
\end{equation} 
where $f_L$ is the Fermi distribution function of the
left reservoir (temperature $T_1=T - \Delta T/2$ and chemical potential
$\mu_1=eV/2$), and $f_R$ is the Fermi distribution function of the right 
reservoir (temperature $T_2=T + \Delta T/2$ and chemical potential 
$\mu_2=-eV/2$). In the linear regime we obtain
\begin{eqnarray} \label{end}
\left[ \begin{array}{c} G \\ B \end{array} \right] =
\frac{e}{2\pi T} \int \left[ \begin{array}{c} -e \\ E \end{array}
\right] t(E) \frac{\partial f}{\partial E} dE\; .
\end{eqnarray}
Here, $G$ and $B$ are the conductance and the thermoelectric coefficient, 
respectively. The thermopower is expressed as $S = -B/G$. The sequel of the 
paper is devoted to the analysis of this expression. We restrict ourselves 
to the case $T \ll \Gamma$, since in the opposite case all structure in the 
function $t(E)$ is washed out by temperature. 

Not too close to the points
$\theta = (2n+1)\pi/2$, $n \in \cal Z$ we obtain
the following asymptotic expressions:
\begin{eqnarray} \label{as}
S = \left\{ \begin{array}{lr}
\displaystyle{\frac{\pi^2 T}{3e\Gamma} (\frac{4\cos\phi}{\cos\theta} -
\frac{\lambda_2}{\lambda_3})} & \epsilon \ll \Gamma \\
\displaystyle{-\frac{\pi^2 T \Gamma}{3\epsilon^2} (\cos\phi \cos\theta
- \frac{\lambda_2}{\lambda_1})} & \vert \epsilon \vert \gg \Gamma
\end{array} \right.\; .
\end{eqnarray}

\begin{figure}
\centerline{\psfig{figure=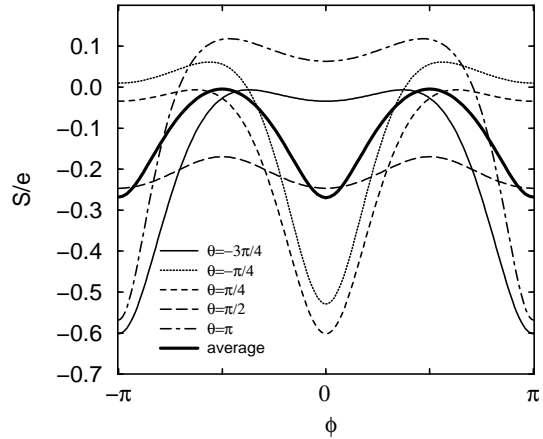,width=8cm}}
\caption{Dependence of the thermopower on the AB flux for $\epsilon
= \Gamma = 5T$ for different values of the phase $\theta$.}
\label{fig2}
\end{figure}

The thermopower shows $\Phi_0$-periodic AB-type oscillations 
as a function of magnetic flux. The oscillations are strong in the sense 
that the thermoelectric effect changes sign as a function of AB
flux. Generally, the shape of these oscillations is anharmonic. In
Fig.~\ref{fig2} we show the thermopower as a function of the AB phase
$\phi$ in the intermediate regime $\epsilon = \Gamma$ for different
values of $\theta$.  

\begin{figure}
\centerline{\vbox{
\psfig{figure=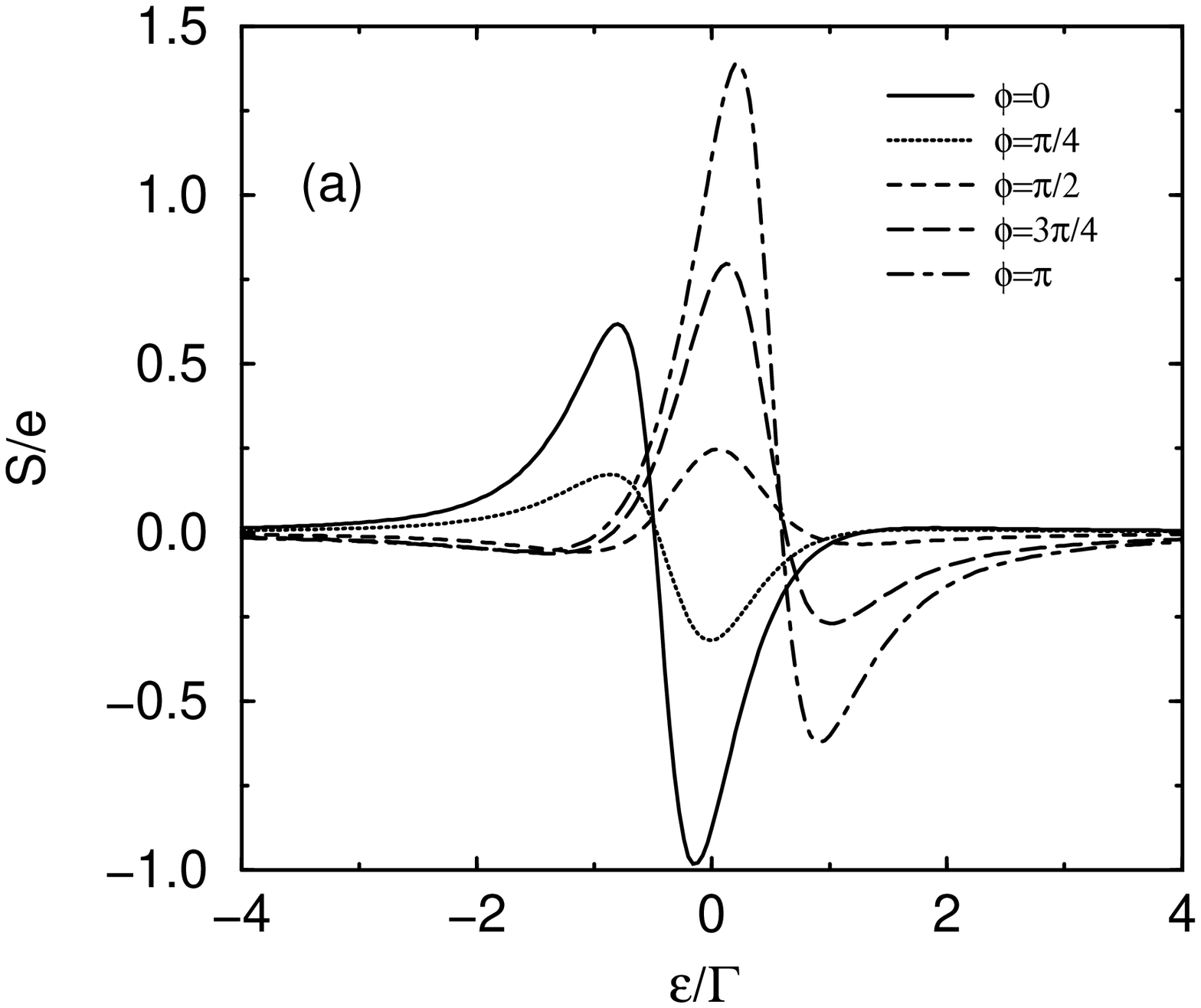,width=8cm}
\psfig{figure=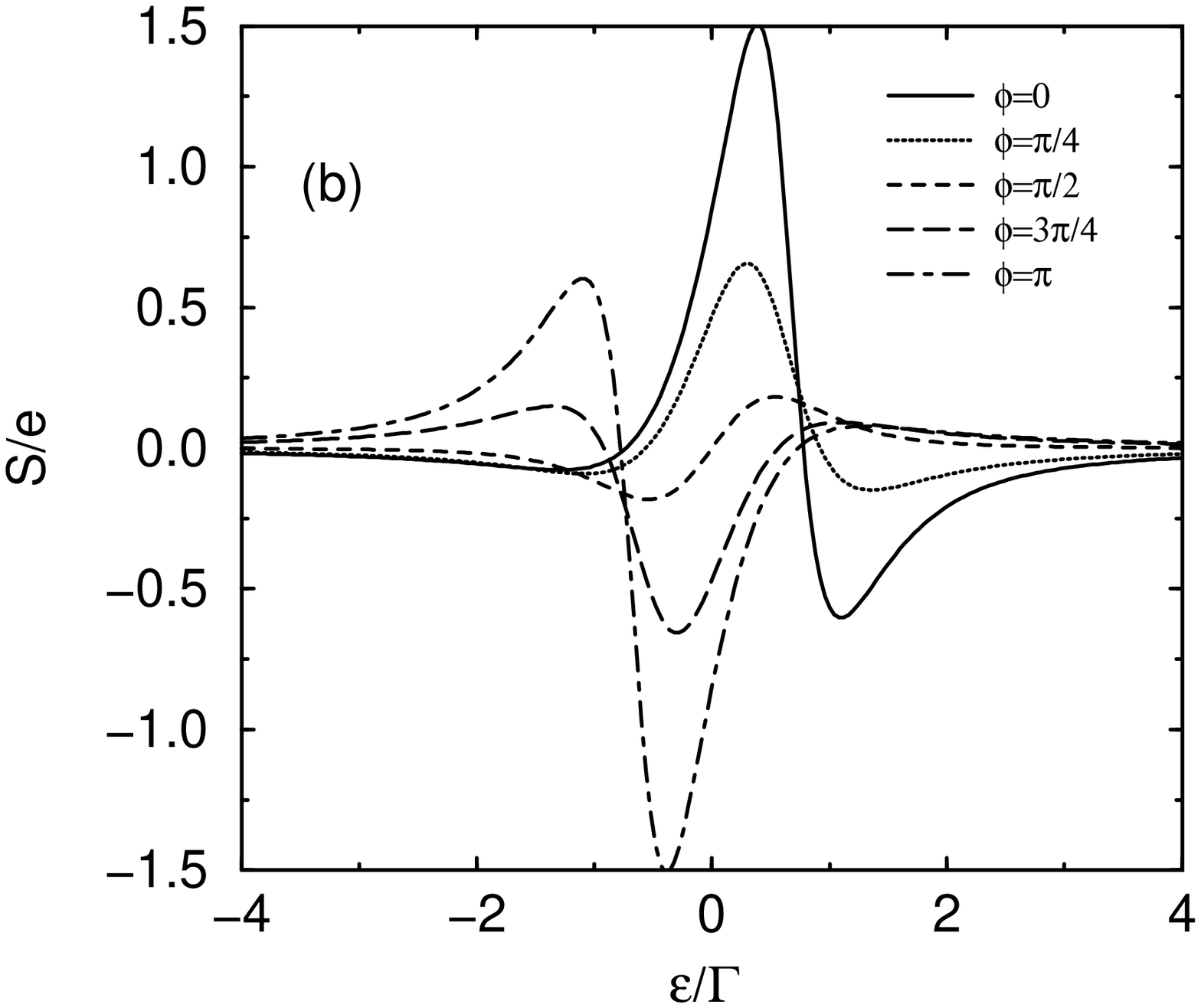,width=8cm}
\psfig{figure=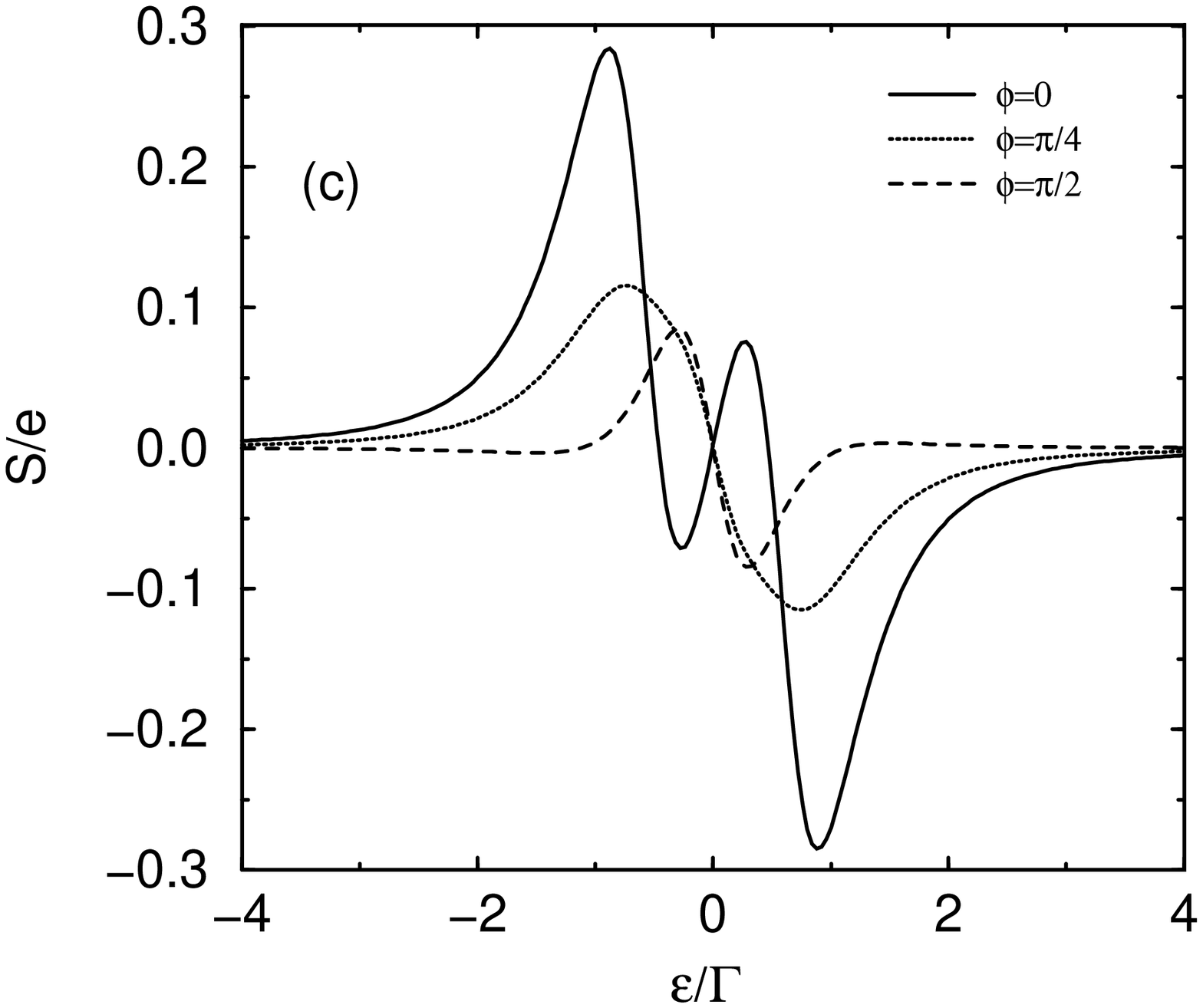,width=8cm}}}
\caption{The gate voltage dependence of the thermopower for $\Gamma =
5T$. (a) $\theta = -3\pi/4$; (b) $\theta = 0$; (c) curves averaged
over $\theta$.}
\label{fig3}
\end{figure}

The gate-voltage dependence of the thermopower is shown in
Fig.~\ref{fig3}. The thermopower shows a characteristic shape with two
maxima around the resonance (see e.g. \cite{Beenakker}) as a function
of the level position $\epsilon$ (or, equivalently, of the gate
voltage). This shape can be easily explained using the
Mott formula \cite{Abr} $S \propto dG/d\epsilon$: since the
conductance exhibits a peak as a function of the gate voltage, its
derivative shows a two-peak structure. 
 
Another important feature is the strong dependence of the shape of the
oscillations and even the sign of the thermopower on the geometric
phase $\theta$, which is controlled by the microscopic details of the
sample and is not known in the given experiment. In this sense we deal
with a typical mesoscopic system: the fluctuations of the thermopower
with respect to the parameter $\theta$ exceed the mean expectation
value. Therefore a direct comparison of the theory with experimental
results is impossible. We suggest, however, two ways to overcome this
difficulty.  

1. The phase $\theta$ can be varied by the gate voltage
(i.e. the level position $\epsilon$). The experiments \cite{Yacoby}
show that the phase is changed by $\pi$ in a narrow window of
gate voltages, so that in this window an explicit dependence
$S(\epsilon)$ is negligible. Hence, in this window one can expect to
observe an unusually strong gate voltage dependence of the
thermopower, originating purely from the dependence $S(\theta)$. 

2. For multi-channel rings or ring ensembles the phase $\theta$ can be
considered as a random quantity, and only the averaged expressions
make sense (see Ref. \onlinecite{Buett2}). The corresponding disorder is
expected to be ``strong'', since a relatively weak variation of the
microscopic structure of the system changes the phase $\theta$
completely. Hence the random variable 
$\theta$ can be considered as uniformly distributed. The corresponding
curves are shown in Fig.~\ref{fig2} and Fig.~\ref{fig3}c. Note that
the averaged thermopower is again a periodic function of the applied
AB flux, but with the period $\Phi_0/2$, i.e. half that of a
given $\theta$. As could be expected, the amplitude of these
oscillations is less than the typical amplitude for arbitrary $\theta$. 

In conclusion, we have considered AB-type oscillations of the
thermopower of a quantum dot embedded in a ring for the case of
non-interacting electrons. We have shown that the oscillations are
strong, and the thermoelectric effect typically changes sign as a
function of AB flux. All details of the oscillations depend
essentially on the microscopic structure of the system. However, ways
to compare the theory with experiment are proposed. 

The regime considered is the most favorable one for AB-type oscillations; 
if the transmission $t$ through the upper arm is small, the amplitude of the 
oscillations is suppressed with $t$ being the corresponding small parameter. 
As a consequence the thermoelectric effect changes sign only in a very narrow
range of parameters, viz. for very low ($\epsilon \ll \Gamma/t$) or very high
($\epsilon \gg \Gamma t$) gate voltages. We expect a different behavior for 
interacting electrons, since the interacting quantum dot can be described by
an Anderson impurity model \cite{Herbert}, and the corresponding physics is
equivalent to Kondo systems. Strong singularities in the behavior of the 
thermopower appear even for an isolated quantum dot \cite{Unp}. However, 
these features are expected to show up only in small and clean dots, whereas 
usually a behavior similar to the one described above will be observed. 

 
The support of the
Deutsche Forschungsgemeinschaft through SFB 195 is gratefully
acknowledged. Y.~M.~B. acknowledges the support of the Alexander von
Humboldt Foundation.


\begin{references}
\bibitem{foot1} We have in mind a quantum well or a quantum dot with one
level in the relevant energy range, i.e., between the chemical potentials
of the left and right reservoirs.
\end{references}
\end{document}